\documentclass[longauth]{aa}
\usepackage{graphicx}
\usepackage{color}
\usepackage{natbib}
\usepackage{txfonts}
\newcommand{\cmtwo}{cm$^{-2}$}

\newcommand{\kms}{km\,s$^{-1}$}       
\newcommand{\vlsr}{$v_{\rm LSR}$}        

\newcommand{\ta}{$T_{\rm A}$}        
\newcommand{\tsys}{$T_{\rm sys}$}

\newcommand{\um}{$\mu$m}                                 
\newcommand{\molh}{H$_{2}$}                              

\newcommand{\water}{H$_{2}$O}

\newcommand{\gapprox}{$\stackrel {>}{_{\sim}}$}   

\newcommand{\about}{$\sim$}                       
\newcommand{\powten}[1]{10$^{#1}$}
\newcommand{\molo}{{\rm O}$_2$}
\newcommand{\otwofull}{${\rm O}_2\,(N_J=1_1-1_0)$}
\newcommand{\roa}{$\rho \, {\rm Oph \, A}$}
\newcommand{\roac}{$\rho \, {\rm Oph \, A \, cloud}$}

\newcommand{\amin}{$^{\prime}$}                   
\newcommand{\asec}{$^{\prime \prime}$}
\newcommand{\adeg}{$^{\circ}$}

\newcommand{\radot}[4]{\mbox{#1$^{\rm h}$#2$^{\rm m}$#3$\stackrel{\rm s}
{_{\bf\cdot}}$#4}}  
\newcommand{\decdms}[3]{\mbox{#1$^{\circ}$#2$^{\prime}$#3$^{\prime \prime}$}}

\newcommand{\adegdot}[2]{\mbox{#1$\stackrel {\circ}{_{\bf \cdot}}$#2}}
\newcommand{\amindot}[2]{\mbox{#1$\stackrel {\prime}{_{\bf \cdot}}$#2}}

\begin{document}

   \title{Molecular oxygen in the $\rho$\,Ophiuchi cloud\thanks{Based 
  on observations with Odin, a Swedish-led satellite project funded jointly by 
  the Swedish National Space Board (SNSB), the Canadian Space Agency (CSA), 
  the National Technology Agency of Finland (Tekes)  and Centre National
  d'Etude Spatiale (CNES). The Swedish Space Corporation has been the industrial 
  prime contractor and also is operating the satellite. $^{\dag}$\,Deceased.}
  }

   \author{     B.\,Larsson\inst{1}       \and
		R.\,Liseau\inst{1}        \and
		L.\,Pagani\inst{2}	  \and
            	P.\,Bergman\inst{3,\,21}  \and
		P.\,Bernath\inst{4}       \and
		N.\,Biver\inst{5}         \and
            	J.H.\,Black\inst{3}    	  \and 
		R.S.\,Booth\inst{3}       \and
	    	V.\,Buat\inst{6}    	  \and
		J.\,Crovisier\inst{5}	  \and
		C.L.\,Curry\inst{4}    	  \and
		M.\,Dahlgren\inst{3}      \and
		P.J.\,Encrenaz\inst{2}    \and 		 
		E.\,Falgarone\inst{7}     \and
                P.A.\,Feldman\inst{8}     \and 
		M.\,Fich\inst{4}    	  \and
                H.G.\,Flor\'{e}n\inst{1}  \and
		M.\,Fredrixon\inst{3}     \and
		U.\,Frisk\inst{9}  	  \and 
		G.F.\,Gahm \inst{1}       \and
		M.\,Gerin\inst{7}    	  \and
		M.\,Hagstr\"om\inst{3}    \and
		J.\,Harju\inst{10}    	  \and		
                T.\,Hasegawa\inst{11}     \and
		{\sc \aa}.\,Hjalmarson\inst{3}\and
                L.E.B.\,Johansson\inst{3} \and
		K.\,Justtanont\inst{1}    \and
		A.\,Klotz\inst{12}	  \and
		E.\,Kyr\"ol\"a\inst{13}	  \and
		S.\,Kwok\inst{11,\,14}    \and
                A.\,Lecacheux\inst{5}     \and
		T.\,Liljestr\"om\inst{15} \and
		E.J.\,Llewellyn\inst{16}  \and
		S.\,Lundin\inst{9}	  \and
		G.\,M\'egie$^{\dag}$\inst{17}\and
		G.F.\,Mitchell\inst{18}   \and
		D.\,Murtagh\inst{19}	  \and
		L.H.\,Nordh\inst{20}      \and
		L.-{\sc \aa}.\,Nyman\inst{21}\and
		M.\,Olberg\inst{3}    	  \and
		A.O.H.\,Olofsson\inst{3}  \and
		G.\,Olofsson\inst{1}      \and
		H.\,Olofsson\inst{1,\,3}  \and
		G.\,Persson\inst{3}	  \and
		R.\,Plume\inst{11}	  \and
		H.\,Rickman\inst{22}	  \and
		I.\,Ristorcelli\inst{12}  \and 
		G.\,Rydbeck\inst{3}	  \and
                Aa.\,Sandqvist\inst{1}    \and
                F.v.\,Sch\'eele\inst{9}   \and
 		G.\,Serra$^{\dag}$\inst{12}\and
		S.\,Torchinsky\inst{23}	  \and
		N.F.\,Tothill\inst{18}    \and
                K.\,Volk\inst{11}         \and
		T.\,Wiklind\inst{3,\,24}  \and
		C.D.\,Wilson\inst{25}     \and	
		A.\,Winnberg\inst{3}	  \and
		G.\,Witt \inst{26} 
	}

   \offprints{B. Larsson}

   \institute{ Stockholm Observatory, AlbaNova University Center, SE-106 91 Stockholm, Sweden \\
   	\email{name@astro.su.se}
    \and       
    	LERMA \& UMR 8112 du CNRS, Observatoire de Paris, 61, Av. de l'Observatoire, 75014 Paris, France
    \and  
        Onsala Space Observatory, SE-439 92, Onsala, Sweden         
    \and         
        Department of Physics, University of Waterloo, Waterloo, ON N2L 3G1, Canada
    \and
	LESIA, Observatoire de Paris, Section de Meudon, 5, Place Jules Janssen, 92195 Meudon Cedex, France
    \and      
        Laboratoire d'Astronomie Spatiale, BP 8, 13376 Marseille Cedex 12, France
    \and
    	LERMA \& UMR 8112 du CNRS, \'Ecole Normale Sup\'erieure, 24 rue Lhomond, 75005 Paris, France
    \and
	Herzberg Institute of Astrophysics, 5071 West Saanich Road, Victoria, BC, V9E 2E7, Canada
    \and       
    	Swedish Space Corporation, P O Box 4207, SE-171 04 Solna, Sweden   
    \and    	
    	Observatory, P.O. Box 14, University of Helsinki, 00014 Helsinki, Finland
    \and    	
    	Department of Physics and Astronomy, University of Calgary, Calgary, ABT 2N 1N4, Canada    
    \and    
    	CESR, 9 Avenue du Colonel Roche, B.P. 4346, F-31029 Toulouse, France
    \and
	Finnish Meteorological Institute, PO Box 503, 00101 Helsinki, Finland
    \and 
	Institute of Astronomy and Astrophysics, Academia Sinica, P.O. Box 23-141, Taipei 106, Taiwan
    \and   
    	Mets\"ahovi Radio Observatory, Helsinki University of Technology, Otakaari 5A, FIN-02150 Espoo, Finland 
    \and
	Department of Physics and Engineering Physics, 116 Science Place, University of Saskatchewan, Saskatoon, SK S7N 5E2, Canada
    \and
	Institut Pierre Simon Laplace, CNRS-Universit\'e Paris 6, 4 place Jussieu, 75252 Paris Cedex 05, France   
    \and      
    	Department of Astronomy and Physics, Saint Mary's University, Halifax, NS, B3H 3C3, Canada
    \and
	Global Environmental Measurements Group, Chalmers Iniversity of Technology, 412 96 G\"oteborg, Sweden
    \and      
        Swedish National Space Board, Box 4006, SE-171 04 Solna, Sweden
    \and
	APEX team, ESO, Santiago,  Casilla 19001, Santiago 19, Chile  
    \and
	Uppsala Astronomical Observatory, Box 515, 751 20 Uppsala, Sweden
    \and
	Canadian Space Agency, St-Hubert, J3Y 8Y9, Qu\'ebec, Canada
    \and
	ESA Space Telescope Division, STScI, 3700 San Martin Drive Baltimore, MD 21218, USA
    \and    	
    	Department of Physics and Astronomy, McMaster University, Hamilton, ON, L8S 4M1, Canada
    \and
	Department of Meteorology, Stockholm University, 106 91 Stockholm, Sweden
    }

\date{Received date: \hspace{5cm}Accepted date:}


\abstract
{Molecular oxygen, \molo, has been expected historically to be an abundant component of the chemical species in molecular
clouds and, as such, an important coolant of the dense interstellar medium.  However, a number of attempts from both ground
and from space have failed to detect \molo\ emission.}
{The work described here uses heterodyne spectroscopy from space to search for molecular oxygen in the interstellar medium.}
{The Odin satellite carries a 1.1\,m sub-millimeter dish and a dedicated 119\,GHz receiver for the ground state line of
\molo. Starting in 2002, the star forming molecular cloud core \roa\ was observed with Odin for 34 days during several
observing runs.}
{We detect a spectral line at \vlsr\,$=+3.5$\,\kms\ with $\Delta v_{\rm FWHM}=1.5$\,\kms, parameters which are also common
to other species associated with \roa. This feature is identified as the O$_2$ ($N_J = 1_1 - 1_0$) transition at
118\,750.343\,MHz.}
 {The abundance of molecular oxygen, relative to \molh\,, is $5 \times 10^{-8}$ averaged over the Odin beam. This abundance
 is consistently lower than previously reported upper limits.}

              
 \maketitle

%

\section{Introduction}

\molo\ is an elusive molecule that has been the target of two recent searches using the Submillimeter Wave Astronomy Satellite
\citep[SWAS,][]{goldsmith2000} and Odin \citep{pagani2003}. Previous attempts included the search for \molo\ with the balloon
experiment Pirog\,8 \citep{olofsson1998} and the search for the rarer isotopomer $^{16}$O$^{18}$O with the {\it Radio
telescope millimetrique} POM-2 \citep{marechal1997} and the Caltech Submillimeter Observatory (CSO) (van Dishoeck, Keene \&
Phillips, private communication) from the ground.  These unsuccessful searches implied very low \molo\ abundances, a highly
intriguing result which led
\citet{bergin2000} to make several suggestions aimed at understanding the low \molo\ abundance. A number of subsequent chemical
models focused on innovative solutions to understand the interstellar oxygen chemistry
\citep[e.g.,][]{charnley2001,roberts2002, spaans2001,viti2001,willacy2002}.

As these models had to rely on upper limits and, as such, were not very well constrained, it seemed likely that an actual
measurement of the \molo\ concentration would increase our understanding of molecular cloud chemistry. For this reason, Odin
has continued the time-consuming task of observing \molo. Here, we report on our renewed efforts to detect \molo\ toward
\roa. First results have been announced previously by \citet{larsson2005} and by \citet{liseau2005}; however, in this paper,
we also account for additional data which were collected in 2006.

\section{Odin observations}
 
The \molo\ observations were done in parallel while Odin was mapping the \roac\ in molecular lines in the submillimeter
bands (submm). These observations were made in August 2002, September 2002, and February 2003 during 20 days of satellite
time (300 orbits), with an additional 200 orbits in February 2006. Due to Earth occultation, only two thirds of the 97\,min
orbit are actually available for astronomy. The total {\it on-source} integration time was 77 hours in 2002, 52\,hr in 2003,
and finally 68\,hr in 2006.

\begin{figure}[t]
  \resizebox{\hsize}{!}{
  \rotatebox{270}{\includegraphics{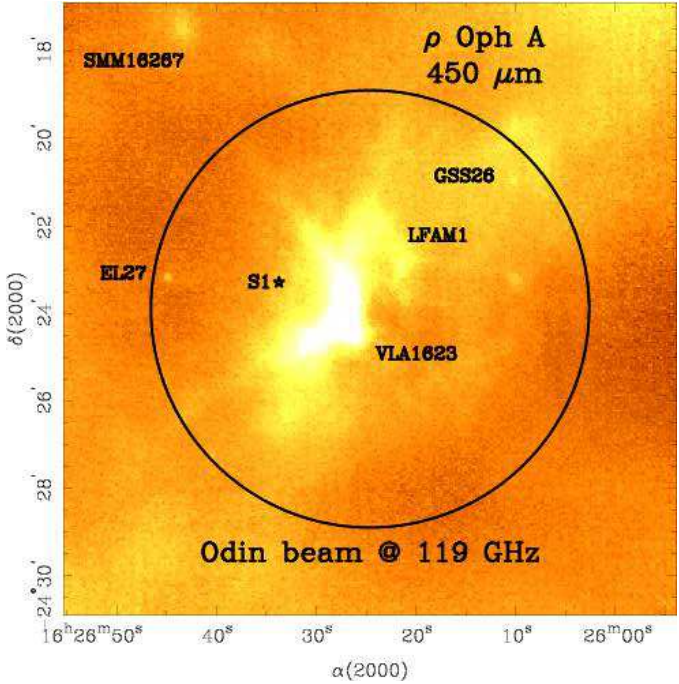}}
                        }
  \caption{The 10\amin\ Odin beam at 119\,GHz, the frequency of \otwofull, is shown as the large circle superposed onto the
	450\,\um\ map by \citet{hargreaves2004} \citep[see also][]{wilson1999}. At the distance of 145\,pc \citep{dezeeuw1999},
	the beam size corresponds to 0.4\,pc. A few discrete sources are labelled. The observations were centered on
	RA\,=\,\radot{16}{26}{24}{6}, Dec\,=\,\decdms{$-24$}{23}{54} (J2000).}
  \label{o2_beam}
\end{figure}

At the frequency of the \otwofull\ line (118\,750.343\,MHz), the beam of the 1.1\,m telescope is 10\amin\
(Fig.\,\ref{o2_beam}).  This large beam implies that the Odin pointing uncertainty ($\le 15$\asec) is entirely negligible
for the \molo\ observations.  In the submm mapping mode, the \molo\ beam moved by less than $1/5$ of its width with respect
to the center position at RA\,=\,\radot{16}{26}{24}{6} and Dec\,=\,\decdms{$-24$}{23}{54} (J2000). The observed off-position,
supposedly free of molecular emission, was 900\asec\ north of these coordinates.

The observing mode was Dicke-switching with the \adegdot{4}{7} sky-beam pointing off by $42$\adeg\
\citep{frisk2003,olberg2003}.  The front- and back-ends were, respectively, the 119\,GHz receiver and a digital
autocorrelator (AC) with resolution of 292\,kHz, channel separation 125\,kHz (0.32\,\kms\ per channel) and bandwidth
100\,MHz (250\,\kms). The 119\,GHz mixer is fixed-tuned to suppress the undesired sideband \citep{frisk2003}. Individual
spectral scans were 5\,s integration each.  The system noise temperature, \tsys, was 650\,K (single sideband) in 2002 and
750\,K in 2003. In 2006, the \tsys\ had increased significantly to 1\,100\ - 1\,300\,K, so these data did not provide any
improvement to the signal-to-noise ratio.

In contrast to the Odin \molo\ search by \citet{pagani2003}, the receiver was no longer phase-locked during the observations
of \roa\ reported here. However, Odin `sees' the Earth's atmosphere for a third of an Odin revolution. Accurate frequency
standards are thus provided by the telluric oxygen lines.

\begin{figure}[t]
  \resizebox{\hsize}{!}{
  \rotatebox{00}{\includegraphics{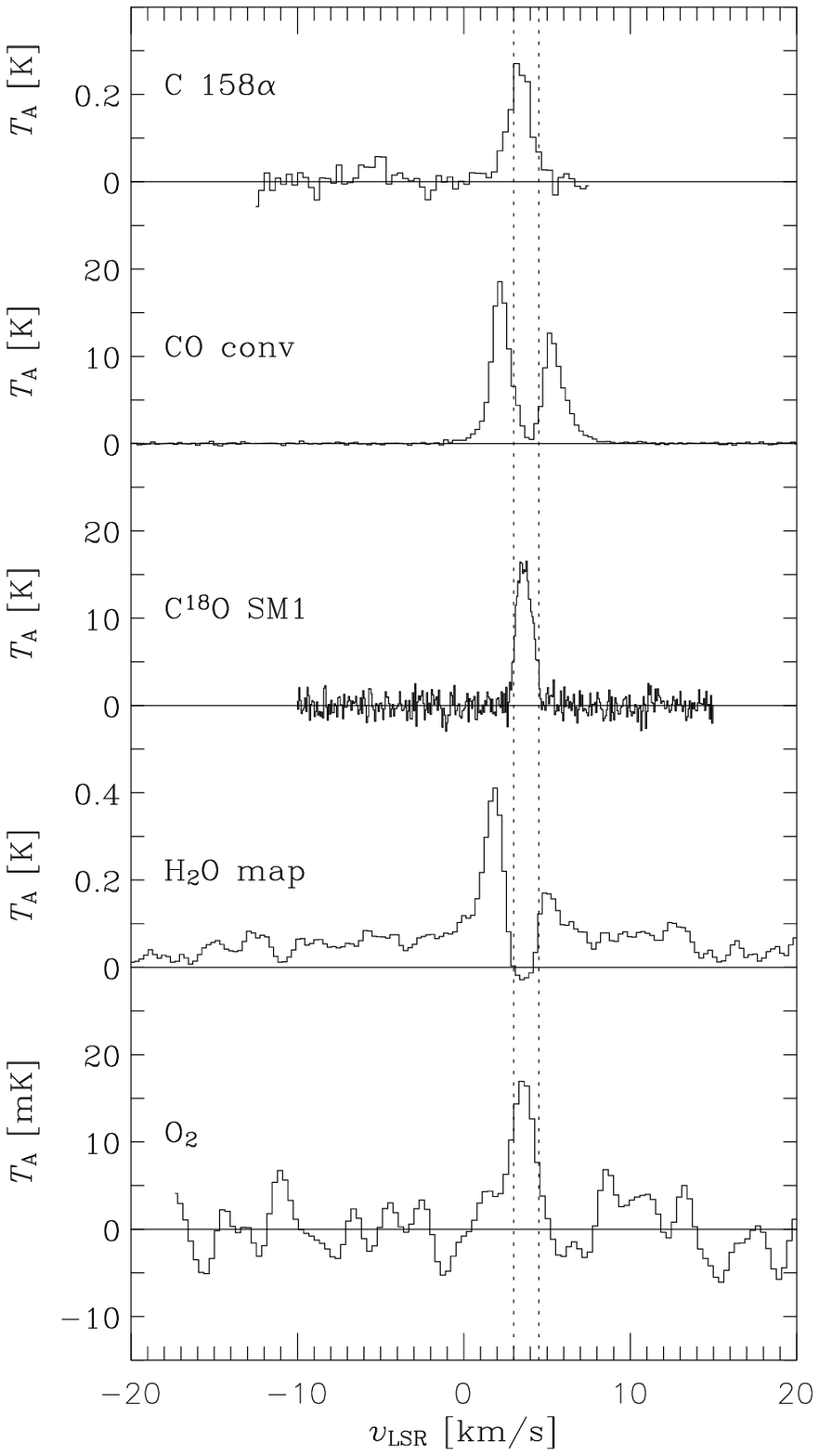}}
                        }
  \caption{The Hanning-smoothed \molo\ line spectrum is shown in the bottom panel and compared to Odin \water\,($1_{1,\,0} - 1_{0,\,1}$) 
       line data \citep {larsson2007}, to C$^{18}$O\,($3-2$) and CO\,($3-2$) from the JCMT-archive, and the C\,$158 \alpha$ recombination
	line from \citet{pankonin1978}. The \water\ spectrum represents the average of the part of the
	\roac\ mapped with Odin (slightly larger than 5\amin\,$\times$\,4\amin).
   	The C$^{18}$O spectrum (15\asec\ beam) refers to a pointed observation toward the Odin center coordinates,
	while the CO spectrum is that of the \amindot{3}{5}\,$\times$\,10\amin\ (at PA=45\adeg) CO\,($3-2$) map
	convolved with the 10\amin\ Odin beam. The displayed spectrum of the C\,$158 \alpha$ recombination line refers to the average of
	4 map positions covering about 10\amin\,$\times$\,10\amin. Vertical dotted lines are shown at $v_{\rm LSR}=+3.0$\,\kms\ and
	$+4.5$\,\kms, respectively.
	}
  \label{o2_PDR}
\end{figure}

\section{The data reduction}

The data reduction method is described in considerably greater detail in Appendix\,A. For the purpose of the following
discussion, it suffices to mention that, because of dramatic frequency variations and receiver instabilities, roughly 50\%
of the \molo\ search data were of too low quality and therefore not included in the final merged spectrum (see
Fig.\,\ref{o2_PDR}). The calibrated spectral data are given in the \ta\ scale. The in-flight main beam efficiency of Odin in
the mm-regime, i.e. at 119\,GHz, is unknown, but is expected not to be much worse than the submm main beam efficiency,
which has been determined as $\eta_{\rm mb}= 0.9$ \citep{frisk2003,hjalmarson2003}.

\begin{figure}[t]
  \resizebox{\hsize}{!}{\rotatebox{00}{\includegraphics{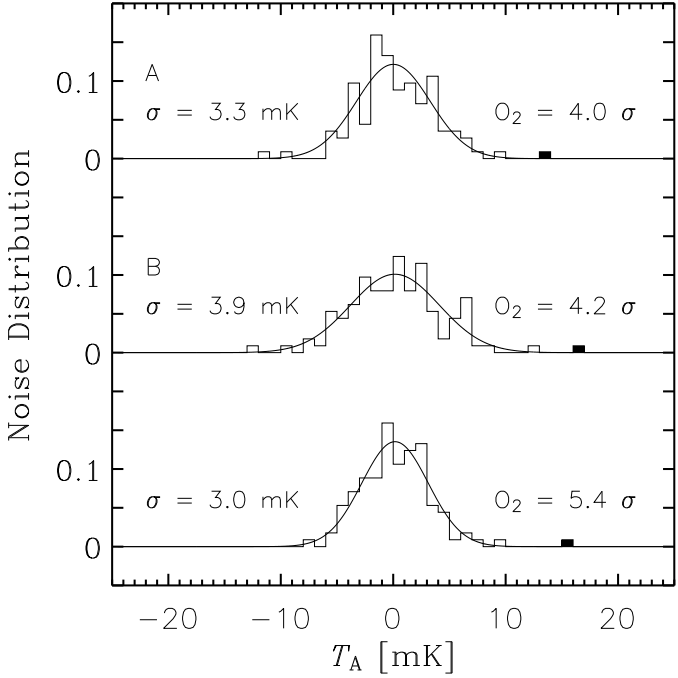}}}
  \caption{The normalized antenna temperature noise distribution. To render the data points statistically independent they
	have been reduced into bins containing 5 channels. As shown in the figure, the statistics of the noise (histograms) are
	all consistent with Gaussian distributions (smooth lines). 
     The letters A and B designate two independent data sets and the entire data set is displayed at the bottom of the figure.
     The respective $4\sigma$ and $5\sigma$ locations of the \molo\ line are shown by the filled-in symbol of the observations.
	}
  \label{o2_prob}
\end{figure}

\begin{table}
\begin{flushleft}
 \caption{\label{o2_parameters} Results for \molo\ observations of \roa}
\begin{tabular}{ll}
  \hline
  \noalign{\smallskip}
Parameter$^{\dagger}$             & Value$^{\ddag}$         \\
  \noalign{\smallskip}
  \hline
  \noalign{\smallskip}
{\it Gaussian fit}                 &                            \\
  \noalign{\smallskip}
\vlsr$_{,\,0}          $\hfill(\kms) & \phantom{1}$ 3.5 \pm 0.3\:(\pm 0.5)^{\#}$ \\
$T_0                   $\hfill(mK)   &            $17.4 \pm 0.1\:(\pm 0.5)^{\#}$ \\
$\Delta v_{\rm FWHM}   $\hfill(\kms) & \phantom{1}$1.5 \pm 0.4\:(\pm 0.5)^{\#}$  \\
$T_{\rm rms}$ \hfill(mK)& \phantom{1}$3.1$            \\
S/N                                & \phantom{1}$5.6$            \\
  \noalign{\smallskip}
  \noalign{\smallskip}
{\it Data 5 channels rebinned}  \\
  \noalign{\smallskip}
$T_0$  \hfill(mK)       & $16.0$   \\
$T_{\rm rms}$ \hfill(mK)& \phantom{1}$2.7$            \\
S/N                                & \phantom{1}$5.9$            \\
$\int\!T {\rm d} v$ \hfill(mK \kms)& $24 \pm 4$                  \\
$N$(\molo) \hfill(\powten{15} \cmtwo)& \phantom{1}1$^{\S}$       \\
  \noalign{\smallskip}
  \hline
  \end{tabular}
\end{flushleft}
Notes to the Table: \\
$^{\dagger}$  Temperatures are in the antenna temperature scale, \ta. $T_0$ are line center and peak temperatures,
respectively. $T_{\rm rms}$ are the fluctuations about the zero-$T$ level over the useful bandwidth, $\Delta v \sim 180$\,\kms. \\
$^{\ddag}$ Gaussian values refer to the AC resolution of 0.3\,\kms\ per channel. \\
$^{\#}$ Including error estimate from the frequency restoration uncertainty. \\
$^{\S}$ For optically thin emission in thermodynamic equilibrium at 30\,K. 
\end{table}

\section{Results}

A part of the \molo\ spectrum is shown in Fig.\,\ref{o2_PDR}, where it is also compared to other spectral lines. At the
top, the observation of a recombination line of carbon is shown \citep{pankonin1978}. Below that, the 
spectrum of the CO\,($3-2$) line from the JCMT archive (James Clerk Maxwell Telescope) shows the spectrally
resolved CO emission averaged over the mapped \roac. The observation in C$^{18}$O\,($3-2$) is toward the center position
of the Odin map and the spectrum of the mapped
\water\,($1_{1,\,0} - 1_{0,\,1}$) line \citep{larsson2007} is displayed below the CO data. 
It is noteworthy that the \molo\ emission feature shares the velocity of the self-absorption feature seen in the optically
thick CO and \water\ lines.

The Odin data over the full spectral range of about 180\,\kms\ are displayed in Fig.\,\ref{o2_spec_parts}. The reduction 
from the initially available 250\,\kms\ to the finally usable
180\,\kms\ bandwidth is due to the fact that some of the channels will be lost when reducing the data (non-matching
alignment; see the figure in Appendix\,A).

In addition to the line at the \molo\ frequency, a feature at three times the rms level can be seen at
the relative Doppler velocity of $+21.0$\,\kms\ (Fig.\,\ref{o2_spec_final}).
The corresponding frequency is 118\,742.0\,MHz, which coincides with the
rest frequency of the $(3_{2,\,2}-2_{1,\,1})$ transition of ethylene oxide of 118\,741.9\,MHz \citep{pan1998}. The apparent
difference of 100\,kHz is within the channel separation of 125\,kHz. Ethylene oxide, c-C$_2$H$_4$O, has been observed in a
number of warm and dense clouds by \citet{ikeda2001}.

\section{Discussion}

\subsection{The validity of the \molo\ line detection}

The \molo\ feature is burried deeply in the Odin data and its extraction requires great care to be taken.
This could initially be seen as a weakness and therefore calls for a rigorous assessment of the validity of the 
claimed detection. In the following, we will present arguments which make the possibility that the observed signal is
of spurious origin highly unlikely.  

\subsubsection{Statistical character of the noise}

Neighbouring spectrometer channels are not independent of each other and the original data have therefore been re-binned 
to the measured width of the line (5 channels, see the Appendix\,A). The resulting distributions of the noise
are shown in Fig.\,\ref{o2_prob}, where the fit to a normalized normal distribution of the combined data set
yields a $\sigma$ of 3.0\,mK. As is evident from Fig.\,\ref{o2_prob}, the statistical significance of the detection 
of the \molo\ line is at the $5.4\sigma$ level.

The probability that the observed line is a noise feature is thus less than $5 \times 10^{-6}$. Furthermore, and very significantly, 
this line is found at the expected Doppler velocity of the \roac\ and since the number of {\it independent} velocity-channels is 
of the order of one hundred, this probability is further reduced to below $5 \times 10^{-8}$. This would apply to a single data set
of observations. This conclusion can be further strengthened by dividing the data into two independent data set (see Appendix\,A)
and performing the same analysis on both sets.

As is evident from Fig.\,\ref{o2_spec_parts}, the line is clearly detected at the correct velocity
in both data sets A and B, at the level of $4.0 \sigma$ and $4.2 \sigma$, respectively (see Fig.\,\ref{o2_prob}). The
corresponding probability for being pure noise is less than $4 \times 10^{-5}$ and $2 \times 10^{-5}$, respectively.
The probability that a pure noise feature will apeare twice at the same channel will therefore be less than $8 \times 10^{-10}$. 


\subsubsection{Systematic effects}

So far, we have considered the nature of the noise only in a statistical sense, but there might of course also be sources of systematic errors. 
Fortunately however, any narrow ($<10$\,\kms) features would be smeared out, when the frequency correction for the drift, both 
from the unlocked receiver and the satellite motion, is made.

Of concern could also be that the integration time spent on the hot load is much less than that spent on the source.
Therefore, as is explained in the Appendix\,A, we approximated \tsys\ with a single value to prevent possible spectral artefacts from the
hot load measurement from entering the observational data.

Another concern could be that the line in the final spectrum is located near the edge of the spectrum. However, more than one third of the
finally used data stems from the 2003 observing campaign. During that period, however, the line was situated in the middle of the spectrometer
band (see the lower right panel in Fig.\,\ref{o2_raw_and_drift}). Therefore, irrespectively of its location in the spectrometer band, the line
is consistently detected at the common (correct LSR-) velocity.

Another problem could be due to leakage from the other sideband. However, to the best of our knowledge there are no spectral lines in
the $2 \times 3.9$\,GHz lower sideband, that could be strong enough to be responsible for the detected \molo\ feature. More important,
however, is the fact that the applied frequency shifts are in the opposite sense in the two bands and that any narrow sideband feature
would become smeared out during the data reduction process.

To summarize, we are also in reasonable control of imaginable systematic effects and, hence, able to counter the most obvious arguments
that would speak in disfavour of the \molo\ detection. Together with the strong statistical arguments, this underlines
the reality of the \molo\ line detected by Odin.

\subsection{Comparison with other data}

\citet{goldsmith2002} have reported a tentative detection of the $(3_3 - 1_2)$\,487\,GHz line of \molo\ toward the \roac\
with SWAS. Both the velocity ($6.0\pm 0.3$\,\kms) and the width ($4.3 \pm 0.7$\,\kms) are incompatible with the Odin
observation of the 119\,GHz line (Table\,\ref{o2_parameters}).

\subsection{Origin of the \molo\ emission}

It is immediately evident from Fig.\,\ref{o2_PDR} that the \molo\ line is similar to the C$^{18}$O\,($3-2$) and the C\,$158 \alpha$
emission and is centered at the position of the self-absorption feature of the optically very thick CO\,($3-2$) and \water\, 
($1_{1,\,0} - 1_{0,\,1}$) lines. Also, the \molo\ line width is consistent with that of the absorption features.  As seen from Earth, a
photon dominated region (PDR) is illuminating the rear side of \roa, with the cool and dense cloud core situated in front of
the PDR \citep{liseau1999}. It is reasonable that the \molo\ emission arises in the molecular core, where nearly a dozen
methanol (CH$_3$OH) lines have previously been observed at \vlsr\ = 3.35\,\kms\ and with $\Delta v_{\rm FWHM}=1.5$\,\kms\
\citep{liseau2003}. These CH$_3$OH transitions were subsequently found to be optically thin. Table\,\ref{o2_parameters}
reveals that the \vlsr\ and $\Delta v_{\rm FWHM}$ for the \molo\ and CH$_3$OH lines are essentially identical, which
suggests that the \molo\ emission is both optically thin (as one would expect) and also originates inside the molecular core
of \roa.

\begin{figure}
  \resizebox{\hsize}{!}{
  \rotatebox{00}{\includegraphics{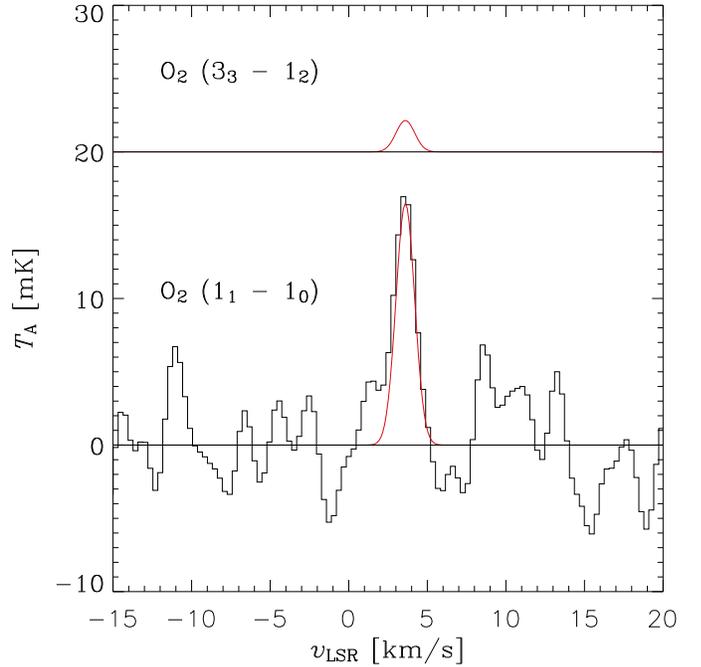}}
                        }
  \caption{A model fit to the \roac, assumed to fill the Odin beam, with a column density of molecular
	hydrogen $N({\rm H}_2)= 2 \times 10^{22}$\,\cmtwo\ and a molecular oxygen abundance of 
	$X({\rm O}_2)= 5 \times 10^{-8}$. The	results for both the 119\,GHz and the 487\,GHz (SWAS and Odin submm
	bands) lines are shown. According to this model, the 487\,GHz line with $T_{\rm peak}=2$\,mK is below
	detectability (see left-hand panel of Fig.\,\ref{o2_noise}).
	}
  \label{o2_model}
\end{figure}

The spectra in Figs.\,\ref{o2_PDR} and \ref{o2_model} show the \molo\ emission integrated over the Odin beam of 10\amin.
Comparison with lines from other species ideally should be done at a comparable angular resolution. \citet{pankonin1978}
observed \roa\ in a carbon recombination line (C\,$158 \alpha$ at 1.6\,GHz) with a beam of \amindot{7}{8}, close in size to
that of Odin. Their beam-averaged \vlsr\ and $\Delta v_{\rm FWHM}$ were $3.15 \pm 0.05$\,\kms\ and $1.43 \pm 0.12$\,\kms,
respectively (see also Fig.\,\ref{o2_PDR}). These values are again very close to the corresponding values of the \molo\ line. 
In contrast to the methanol emission, this recombination radiation certainly originates in the PDR.

Based on the available observational evidence, it seems that no convincing conclusions regarding the location of the \molo\
emission region can be drawn. The resolution to this problem would require the measurement of several \molo\ transitions, a
task which will likely be accomplished with the heterodyne instrument HIFI aboard Herschel\footnote{
\texttt{http://www.sron.nl/divisions/lea/hifi/} }, to be launched into space in the 2008 time frame.

\subsection{The \molo\ abundance}

Of interest is the interpretation of these Odin observations in terms of the abundance of molecular oxygen, $X({\rm O}_2)
\equiv \int\!n({\rm O}_2)\,dz/\!\!\int\!n({\rm H}_2)\,dz = N({\rm O}_2)/N({\rm H}_2)$.  Assuming optically thin line
emission in thermodynamic equilibrium
\citep[for~a~discussion,~see][]{liseau2005}, the observed line intensity implies a column density
$N({\rm O}_2) = 1 \times 10^{15}$\,\cmtwo\ (see Table\,\ref{o2_parameters}). This calculation assumes a temperature of
30\,K, a value which has been consistently obtained for both the gas \citep[e.g.,][]{loren1990} and the dust
\citep{ristorcelli2007}. The latter authors obtained $50^{\prime} \times 40^{\prime}$ maps in four FIR/submm wavebands,
from 200\,\um\ to 580\,\um, using the balloon experiment PRONAOS \citep{pajot2006}. The data provide constraints on the dust
temperature and the spectral emissivity, which lead to maps of the dust optical depth. Assuming a total mass absorption
coefficient of $5 \times 10^{-3}$\,cm$^2$ g$^{-1}$ at 1300\,\um\, with frequency dependence $\propto \nu^2$, the hydrogen
column density is estimated to be $N({\rm H}_2)=1 \times 10^{22}$\,\cmtwo. This column density is at the low end of recent
estimates, which were based on gas tracers \citep{pagani2003,kulesa2005}. Therefore, on the 10\amin\ scale ($10^{18}$\,cm)
of the Odin observations, an average value of $N({\rm H}_2) = 2 \times 10^{22}$\,\cmtwo\ is adopted here.

In conclusion, and acknowledging a factor of two uncertainty at least, the \molo\ abundance in the \roac\ is $X({\rm O}_2)=
5 \times 10^{-8}$ (Fig.\,\ref{o2_model}). With regard to model predictions (see Sect.\,1), these results are consistent with
either very young material of only a few times \powten{5}\,yr, which, given the ages of the stellar population, is unlikely,
or more evolved gas at an age of some Myr to ten Myr, which requires careful `tuning' in terms of selective desorption
and/or specific grain-surface reactions, and thus appears unlikely, too. More distinct statements would need to include
other chemical species in the analysis, which will be the subject of a forthcoming paper.

\section{Conclusions}

Below, we briefly summarize our main conclusions. 

\begin{itemize}
\item[$\bullet$] Dedicated observations with Odin of the \roac\ at 119\,GHz have revealed a spectral feature at the frequency of the
\molo\ ($N_J = 1_1 - 1_0$) transition.
\item[$\bullet$] Both the center velocity and the width of this feature are consistent with other optically thin emission lines from \roa. 
\item[$\bullet$] Analyzing the statistical properties of the noise essentially rules out the possibility that the \molo\ feature is entirely due
to noise in the observations.
\item[$\bullet$] Specific reduction procedures have been applied to the data, which also helped to suppress any possible systematic effects.
\item[$\bullet$] The significance of the line is more than $5\sigma$ and the beam-averaged \molo\ abundance in the \roac\ is $5 \times 10^{-8}$.
\end{itemize}

\begin{acknowledgements} We wish to express our gratitude to the teams at the Odin operation centers of the
Swedish Space Corporation for their skillful and dedicated work. The very careful reading and constructive criticism
of the manuscript  by the referee, M.\,Gu\'elin, is highly appreciated.
\end{acknowledgements} 

%
%

\appendix

\section{Data reduction}

\begin{figure*}
\begin{center}
  \resizebox{\hsize}{!}{
  \rotatebox{90}{\includegraphics{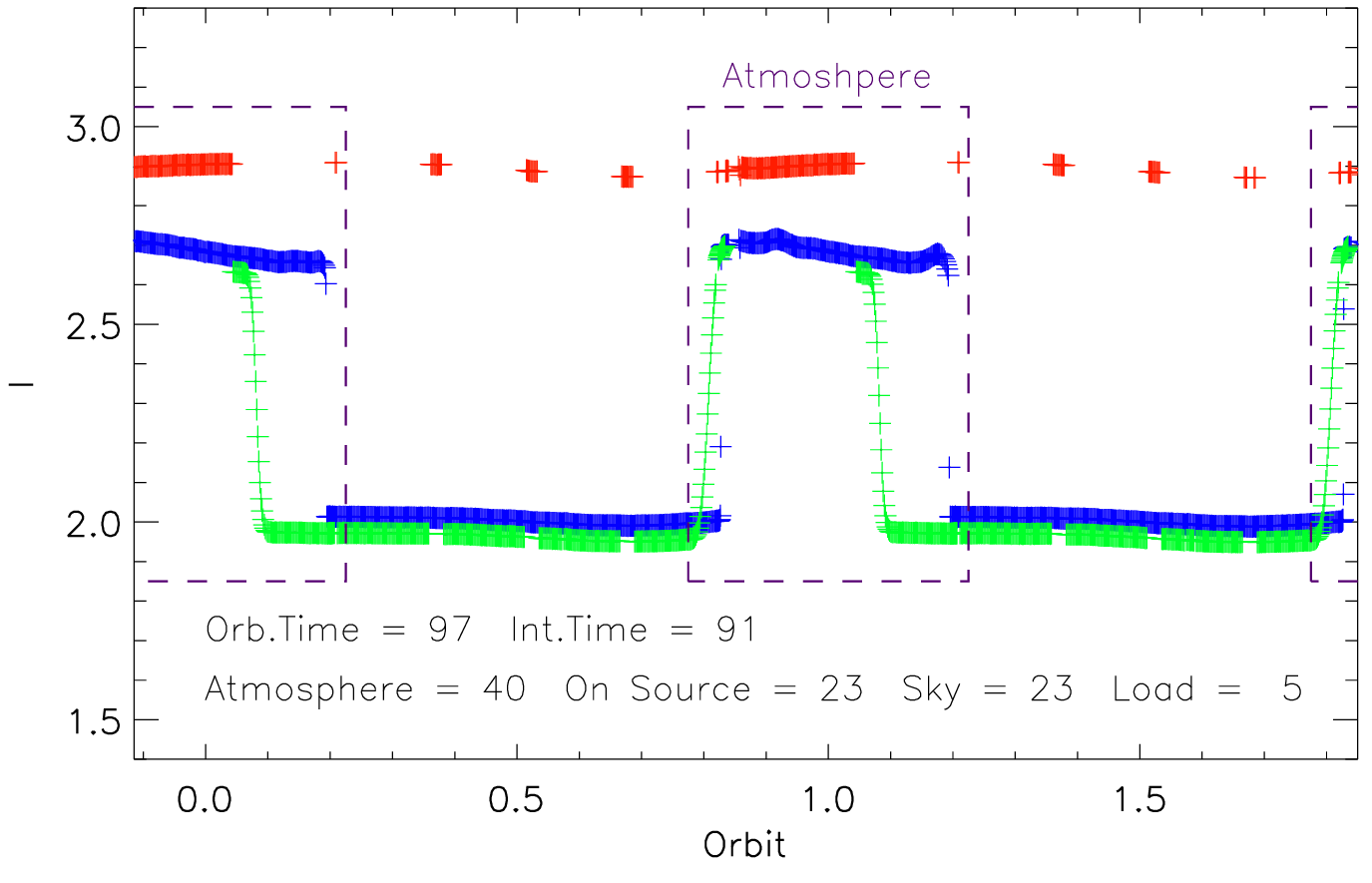}}
  \rotatebox{90}{\includegraphics{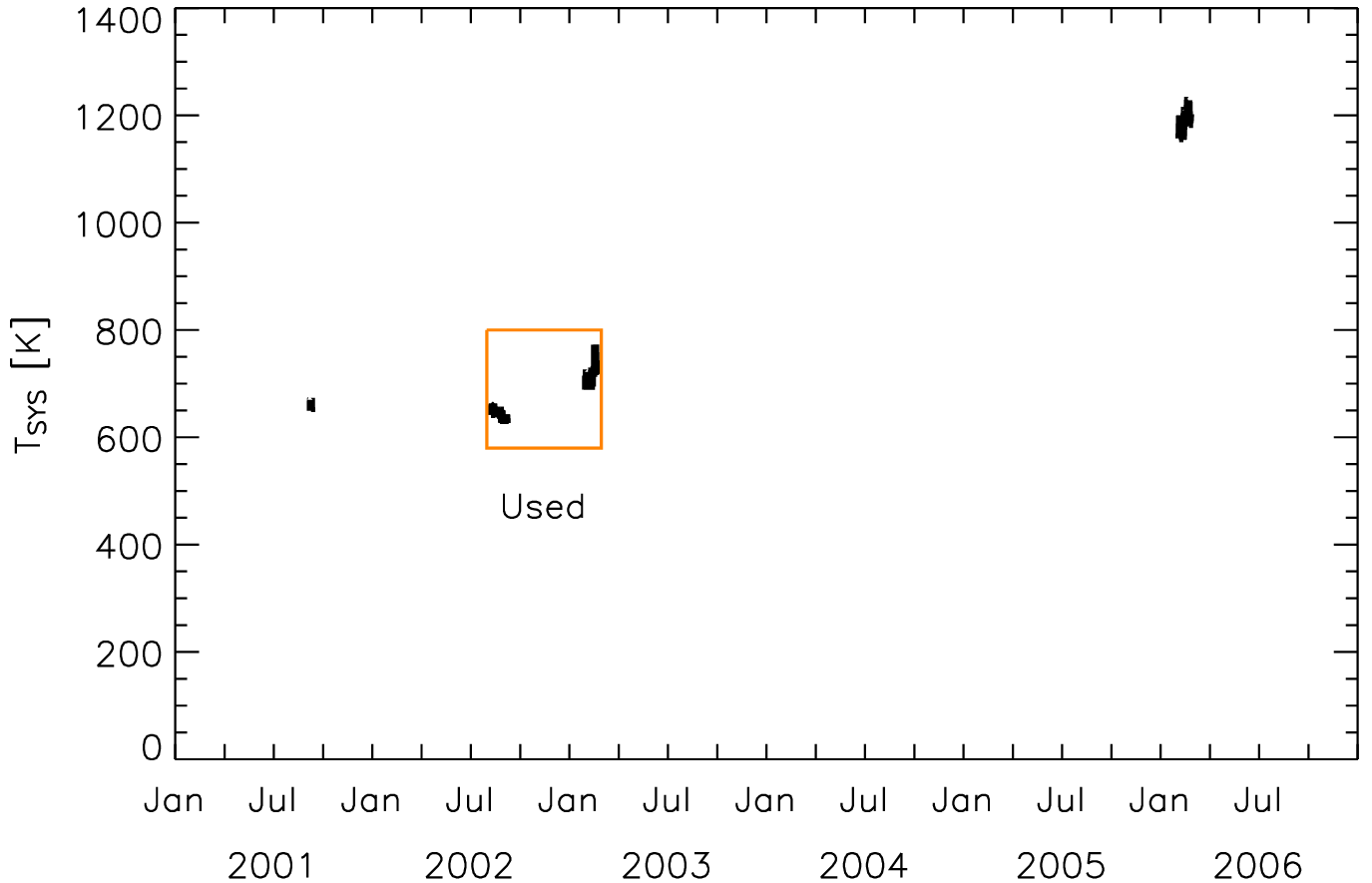}}
                       }
  \caption{Selection procedure of usable data to be analysed - step\,1.
	{\bf Left:} The total power (in instrument units) during an orbit is shown. The colour coding is as follows:
		red = internal hot load (uppermost curve), blue = on source (intermediate) and green = sky measurement (lowest).
		The dashed rectangular boxes show the times of telluric atmosphere observations.
	{\bf Right:} System noise temperature, \tsys, during the four observing runs in 2001, 2002, 2003 and 2006,
		respectively. The red rectangular box identifies the selected data; 2001 data are of lower resolution and
		were, as such, excluded from the present analysis \citep[but~see:][]{pagani2003}.
	}
  \label{o2_observed}
\end{center}
\end{figure*}

\begin{figure*}
\begin{center}
  \resizebox{\hsize}{!}{ \rotatebox{90}{\includegraphics{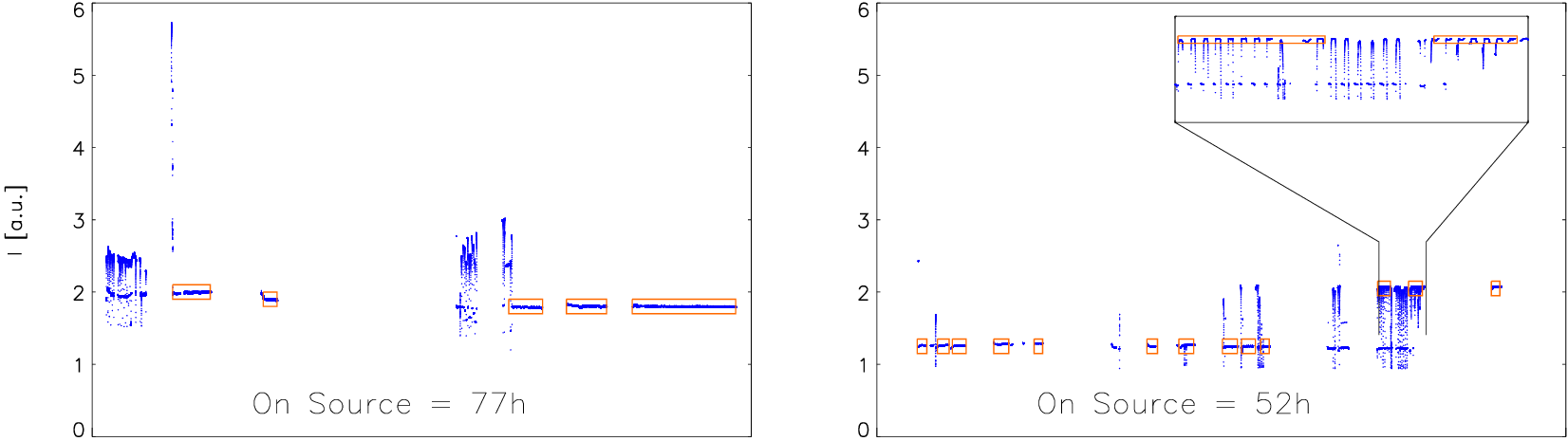}} }
  \resizebox{\hsize}{!}{ \rotatebox{90}{\includegraphics{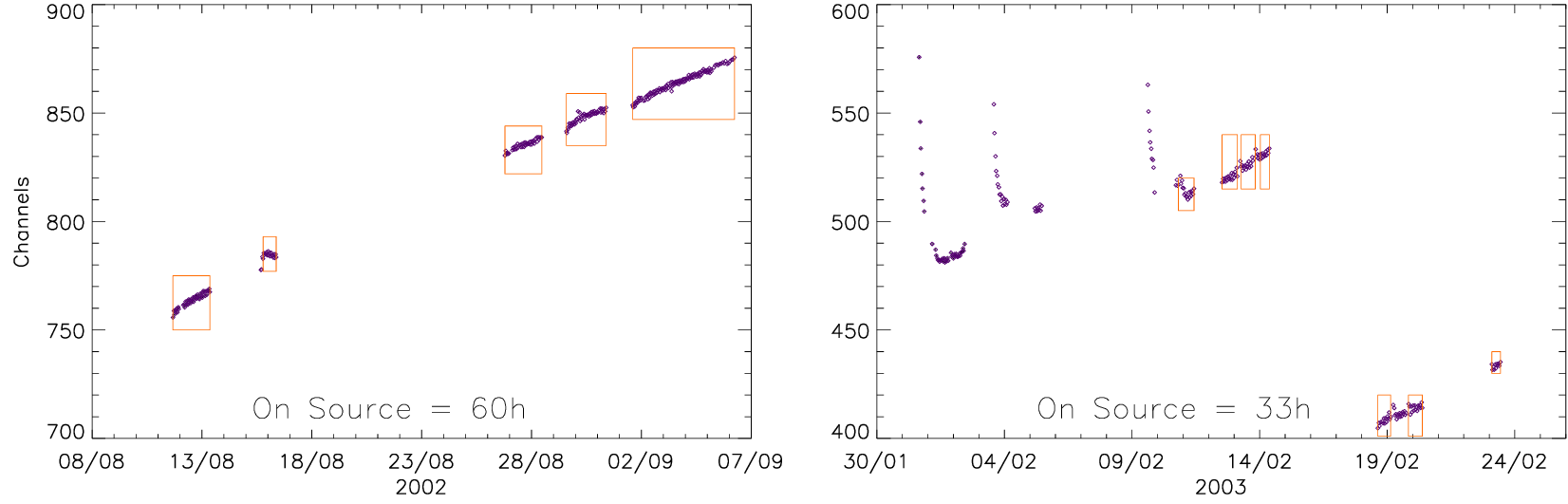}} }
  \caption{Selection procedure of usable data to be analysed. Displayed are data for 2002 and 2003.
	{\bf Top panels:} Step\,2. On-source raw total power data (in instrumental units). Red boxes include used data.
	{\bf Bottom panels:} Step\,3. Central spectrometer channel as function of the time for the all the data selected
	from the above criteria. Notice the change in scale, 
	showing the long time drifts of the 119\,GHz receiver. The jump in 2003 is due to the deliberate resetting
	of the spectrometer to keep the line near the center.
	Measurements of the frequency drift of the receiver were done for the data included in the boxes.
	}
  \label{o2_raw_and_drift}
\end{center}
\end{figure*}

\begin{figure*}
\begin{center}
  \resizebox{\hsize}{!}{ \rotatebox{90}{\includegraphics{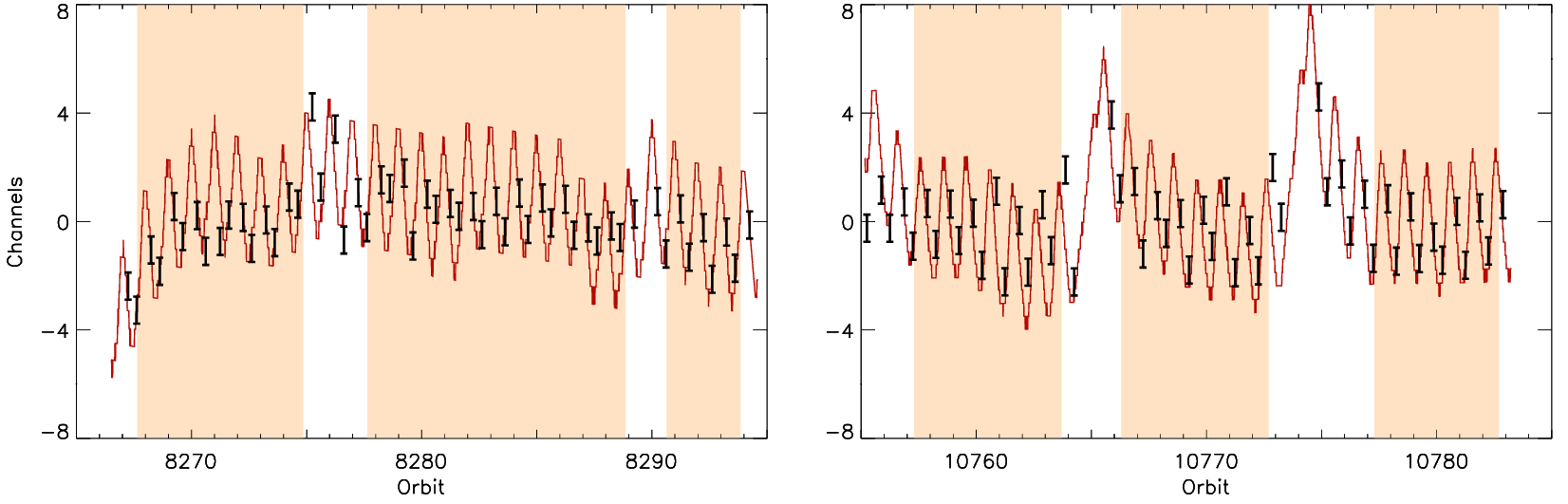}} }
  \resizebox{\hsize}{!}{ \rotatebox{90}{\includegraphics{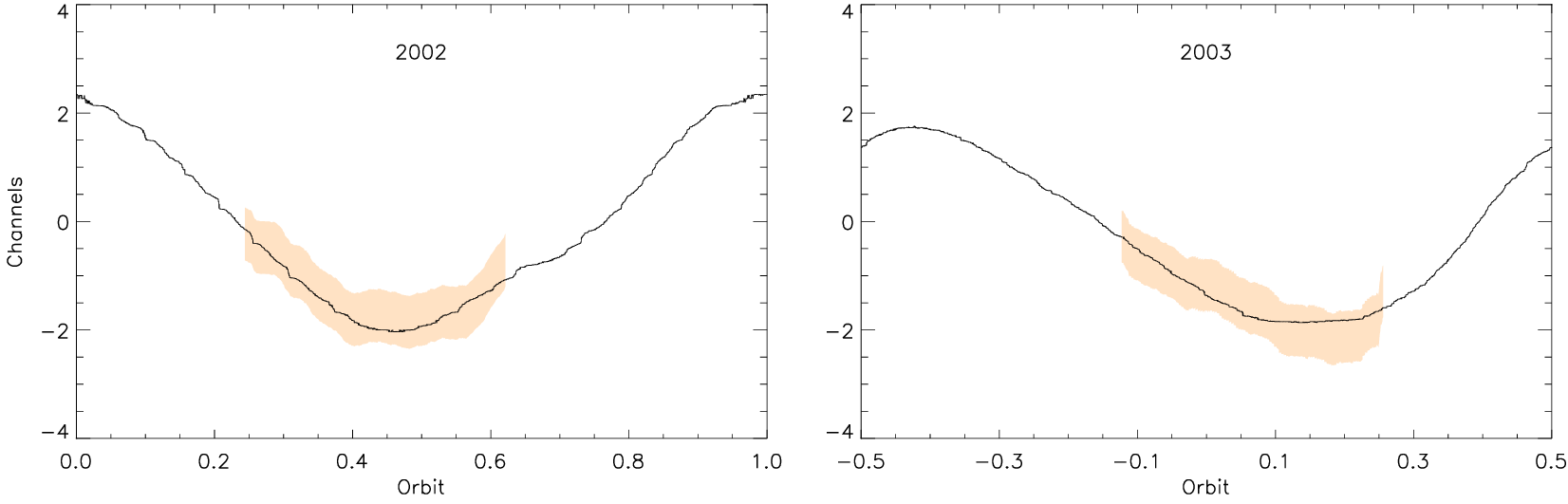}} }
  \caption{Frequency drift correlation with thermistor temperature (at the DRO).
	{\bf Upper panels:} Measurement of the center channel for the 119\,GHz oxygen line, when
		entering and leaving the Earth's atmosphere, respectively. The red full-drawn line
		designates the inverse DRO-temperature, scaled in absolute level to the distribution of center channels,
		the assigned `error bars' of which are $\pm 0.5$ channel widths. The shaded areas mark the selected data.
	{\bf Lower panels:} As above, but for a single orbit. The shaded areas indicate the measurements of the center channel of
		the oxygen line for the entire passage of the Earth's atmosphere. The hight of the shaded areas correspond to the error
		estimates of $\pm 0.5$ channel widths and the full-drawn line is the inverse DRO-temperature variation over
		the whole orbit.
	}
  \label{o2_drift_corr}
\end{center}
\end{figure*}

\begin{figure*}
\begin{center}
  \resizebox{\hsize}{!}{ \rotatebox{90}{\includegraphics{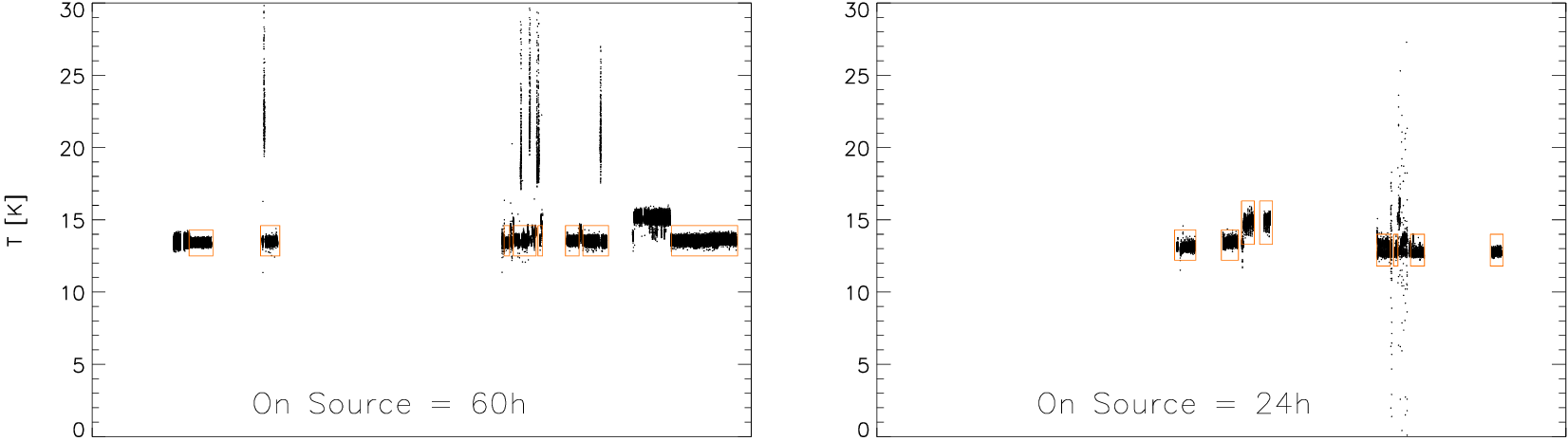}} }
  \resizebox{\hsize}{!}{ \rotatebox{90}{\includegraphics{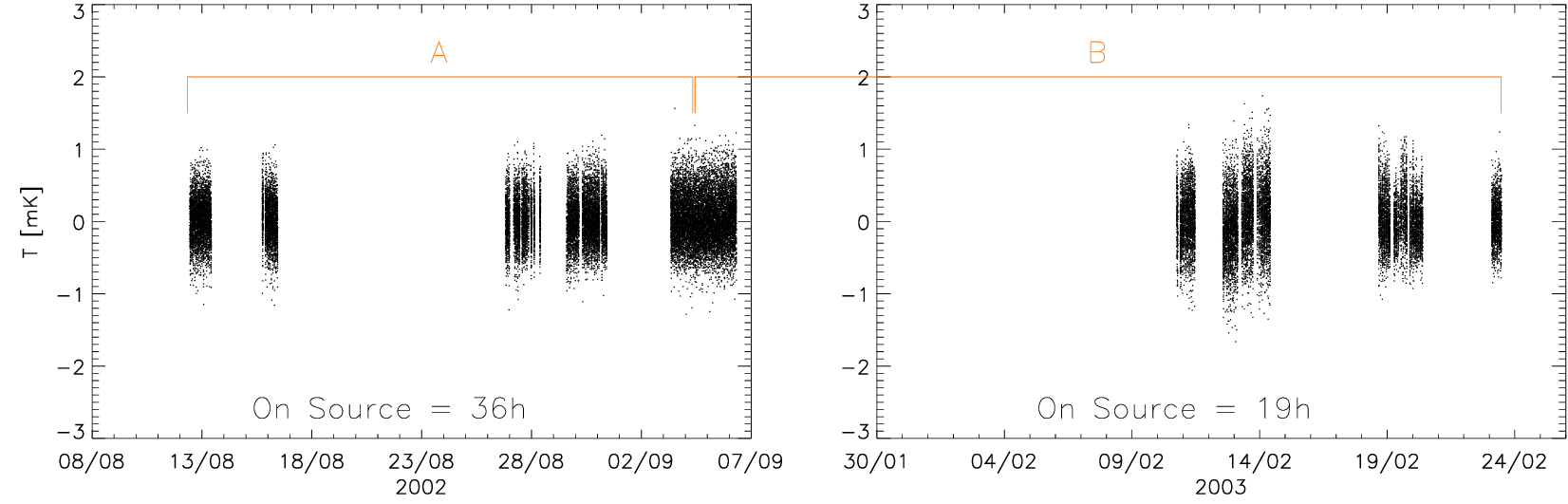}} }
  \caption{Selection procedure of usable data to be analysed - step\,4.
	Top panel: On source calibrated data (mean power)
	Bottom panel: Final selected data after baseline removal.
	The bars, A and B, marks the division of the data set into two equal parts.
	}
  \label{o2_reduce}
\end{center}
\end{figure*}

Except for a short period in the beginning of the Odin mission, the 119\,GHz receiver has not been phase-locked. This has the
consequence of both drifts in LO-frequency and stochastic variations in the amplifier gain. It is of imperative importance,
therefore, that extreme care is exercised when reducing data which have been collected with this receiver and which are
expected to be extremely weak. This is so in order not to create artificial signals nor to suppress faint real ones.

Odin has observed the \roac\ at 119\,GHz on four different occasions, viz. in 2001 \citep[see:][]{pagani2003}, in August
2002, in February 2003 and, finally, in February 2006. During the early observations (2001 and 2002) the system noise
temperature was relatively stable, i.e. \tsys\,\about\,650\,K (Right panel in Fig.\,\ref{o2_observed}). When `left alone'
with the power on, the unlocked system is drifting towards lower frequencies in a nearly linear fashion. Eventually, the
end of the receiver-band is approached, resulting in the increase of \tsys. Consequently, in 2003, the system temperature
had increased to 750\,K. Even worse was the system behaving in 2006, when we attempted to support our earlier observations of
the \molo\ line by additional observations. At that stage, \tsys\ had risen to 1200\,K, making these data essentially of nil
value. In addition, the 2001 data were obtained at lower resolution and, therefore, the results presented in this paper are
based only on the data collected during the 2002 and 2003 observing campaigns.

\subsection{Selective data reduction}

Because of the instability of the 119\,GHz receiver, data collected with this instrument are of highly uneven quality. For
this reason, a sizable fraction of the available data had to be abandoned. In the following, we identify and quantify the
criteria which we applied for the selection of the useful data.

\subsubsection{Gain stability}

When observing with Odin, an astronomical object is blocked by the Earth during \gapprox\,1/3 of the orbit (left panel of
Fig.\,\ref{o2_observed}). The observations were made in Dicke-switching mode, where the receiver alternates between the
source and one of the sky beams and the internal hot load \citep[Sect.\,2.~and][]{frisk2003,olberg2003}. For each orbit, the
total integration time toward the source is therefore limited to 23 minutes. For the entire observing period, the total
integration time for \roa\ was 136 hours, partitioned as 77\,hr in 2002 and 59\,hr in 2003, respectively.

In the top panels of Fig.\,\ref{o2_raw_and_drift}, the uncalibrated on-source data are shown. From the figure, it is evident
that the gain is not stable, but varies violently. Only data, for which the gain was stable or varied only slowly, were
selected for further reduction. This reduced the integration times to 60\,hr for 2002 and 33\,hr for 2003.

\subsubsection{Frequency drift correction}

Odin `sees' the Earth's atmosphere for a third of an Odin revolution (left panel Fig.\,\ref{o2_observed}).  Accurate
frequency standards are thus provided by the telluric oxygen lines, viz. the $N_J=1_1 - 1_0$ line of \molo\ and of its
isotopic variant $^{16}{\rm O}^{18}{\rm O}$.

As discussed above, the frequency drift of the 119\,GHz receiver with time is relatively linear - as long as the receiver
stays powered on.  This was essentially the case in the beginning of the Odin mission, when the 119\,GHz receiver was almost
always turned on. However, as outlined by \citet{pagani2003}, it became clear relatively soon that \molo\ observations
generally resulted in non-detections. Parallel 119\,GHz observations became therefore essentially cancelled, with the aim to
keep the \molo\ line within the receiver band as long as possible.

Henceforth, the \molo\ receiver was powered on only for times when a dedicated oxygen search was performed. However,
initially (during the first few orbits) when the receiver is turned on, the drift is fast and highly non-linear (lower
panels Fig.\,\ref{o2_raw_and_drift}). As described by \citet{larsson2003} for the 572\,GHz receiver, this drift is
temperature dependent.

A thermistor is placed at the Dielectric Resonator Oscillator (DRO). It has been established empirically that the frequency
drift of the unlocked system correlates well with the temperature measured there (upper panel
Fig.\,\ref{o2_drift_corr}). During any given single orbit, the frequency variations should be small, however. This, in fact,
has been confirmed by means of monitoring the atmospheric portion of the orbit, as is shown in the lower panels of
Fig.\,\ref{o2_drift_corr}.

With this good understanding of the behavior of the receiver and spectrum analyzer, the frequencies during the actual
observations of the astronomical sources can be restored to high precision. We iterate here that the reliability of the
reduction method had been demonstrated earlier by the successful reconstruction of the HC$_3$N ($J=13-12$) line at
118\,270.7\,MHz in a number of sources \citep[e.g.,~for~DR\,21,~see][]{hjalmarson2005}. Although we had shown that our
reduction method works well and reliably, none of the data with high drift rates (in the beginning of the observing periods
in 2003) have, after all, been used in the present analysis (see boxes in the right lower panel of
Fig.\,\ref{o2_raw_and_drift}).

\subsection{Final data selection}

\subsubsection{Baseline removal}

The observations were done in Dicke-switching mode, alternating between the 10\amin\ main-beam, one of the \adegdot{4}{7}
sky-beams (pointing off by $42$\adeg), and the internal hot load. The calibrated signal, in the antenna temperature scale,
is therefore given by

\begin{equation}
T_{\rm A} = \frac {I_{\rm On} - I_{\rm Sky} }{ I_{\rm Load} - I_{\rm Sky} } = \frac { I_{\rm On} - I_{\rm Sky} } { I_{\rm Sky} } \times T_{\rm sys}\,\,.
\end{equation}

On the basis of the observation of other optically thin lines, the \molo\ line toward \roa\ is expected to be rather
narrow. In order not to include possible artificial spectral features from the hot-load an average value around the line was
used for the system noise temperature and adopted for the entire band, i.e.

\begin{equation}
T_{\rm sys} = \left < \frac { I_{\rm Sky} }{ I_{\rm Load} - I_{\rm Sky} } \right >\,\,.
\end{equation}

As back-end a digital autocorrelator (AC) was used. Using all 8 available correlator-chips gives a single band with
resolution of 292\,kHz (with channel separation = 125\,kHz or 0.32\,\kms) and bandwidth 100\,MHz (corresponding to
250\,\kms). The main beam and the sky-beams do not match perfectly, implying that the calibration results in a curvature
over the band. An off-position, supposedly free of molecular emission, and 900\asec\ north of \roa\ was also observed and
these data could have been used to correct for the baseline residual. However, since the time spent on the off-position was
much less than that spent on the on-position, the noise level in the difference spectrum would have been dominated by the
observation of the off-position.

We resorted therefore to another method for the baseline removal. During the orbit about the Earth, the central
Doppler-velocity of the line changes within the interval $-7$\,\kms\ to $+ 7$\,\kms. Averaging over one or more orbits will
therefore cause any narrow line to be smeared out. So for every observing period, a ``super baseline'' was constructed (and
smoothed) and removed from every individual spectral scan during that period (bottom panels Fig.\,\ref{o2_reduce}). This
resulted finally in a useful integration time of 36\,hr in 2002 and 19\,hr in 2003 (47\% and 32\% of total on-time,
respectively; see Sect.\,A.1.1).

\begin{figure*}
  \resizebox{\hsize}{!}{ 
	\rotatebox{00}{\includegraphics{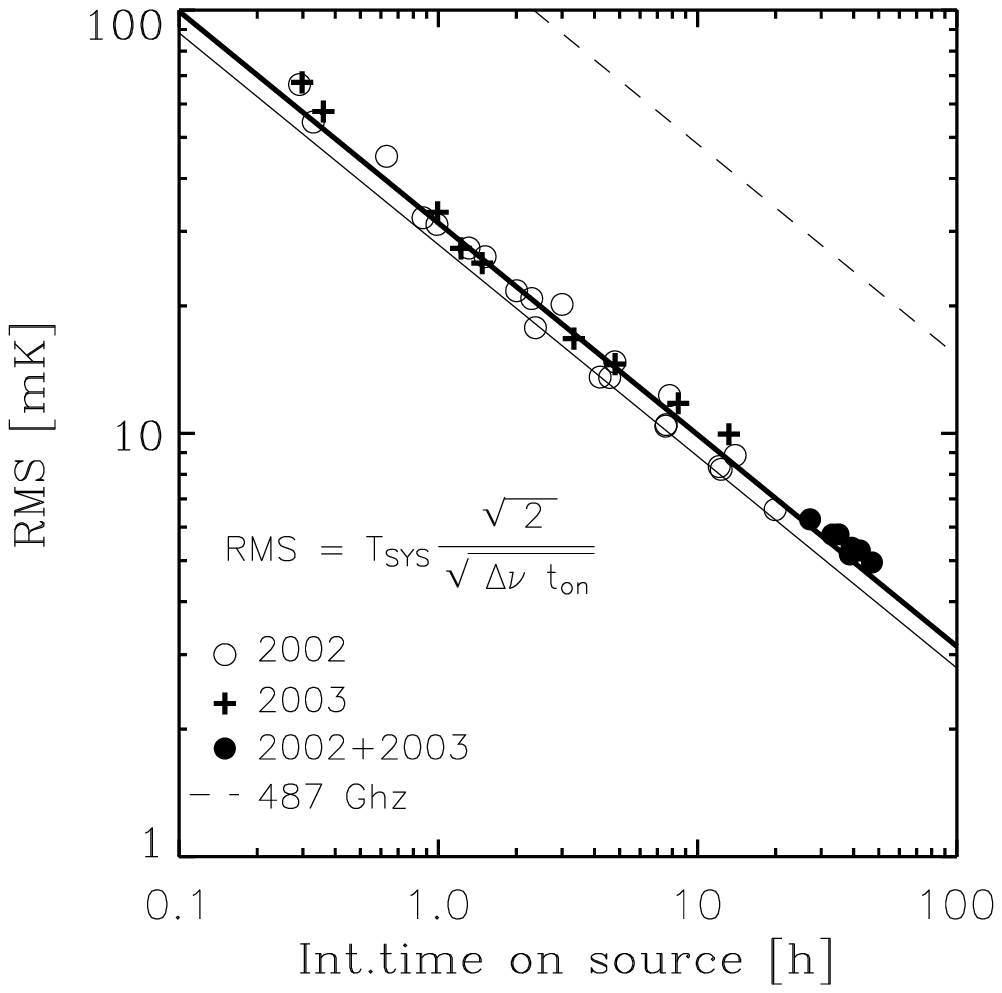}}
	\rotatebox{00}{\includegraphics{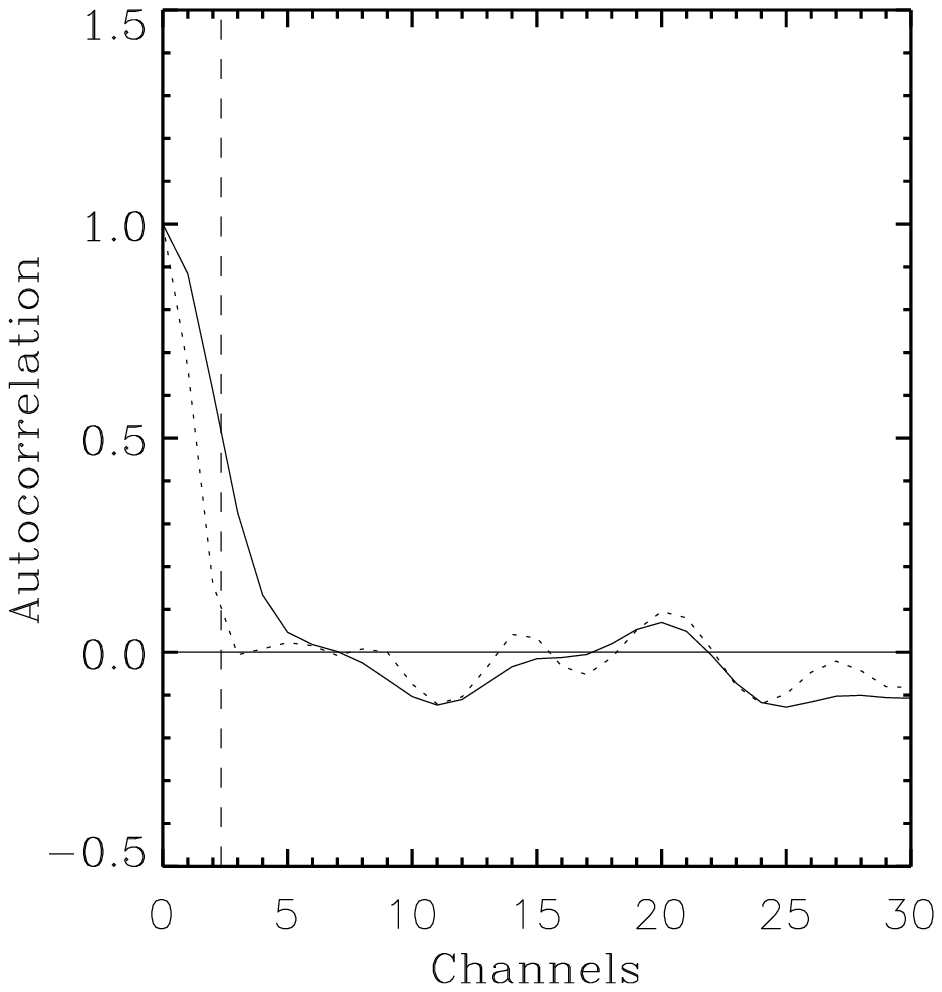}}
	}
  \caption{ {\bf Left:} Noise temperature as a function of integration time ($t_{on}$): 
		Thin and thick lines are for ideal rms noise for \tsys\ of 640\,K (2002) and 720\,K (2003),
 		respectively and a resolution of 292 kHz. The dashed line refers to the expected rms-level
		of the 487\,GHz line of \molo\ (see the text and Fig.\,\ref{o2_model}).
	{\bf Right:} Autocorrelation of the finally reduced spectrum.
	Dotted line: before using a Hanning filter.
	Full line: after using a 3-channel-wide Hanning filter.
	Dashed line: the 292 kHz spectral resolution.}
  \label{o2_noise}
\end{figure*}

\begin{figure*}
  \resizebox{\hsize}{!}{ 
	\rotatebox{00}{\includegraphics{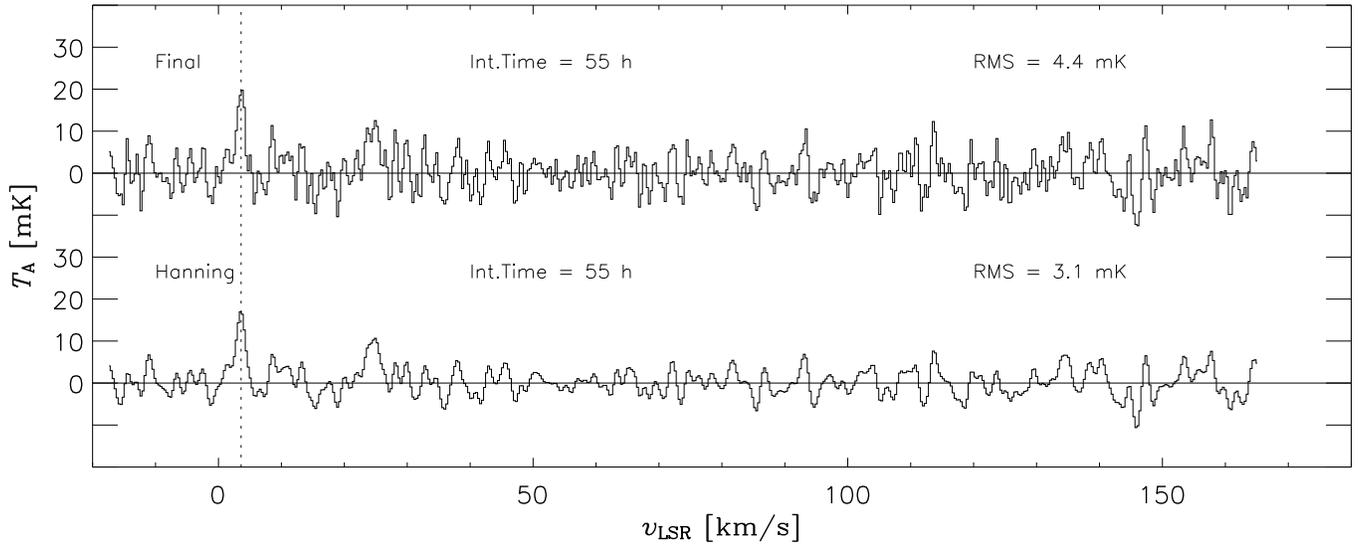}} 
	}
  \caption{ {\it Final}: The final spectrum (\ta\ in mK versus \vlsr\ in \kms). Besides the \molo\ line near zero LSR-velocity,
		also another feature is discernable at about $+20$\,\kms\ (see the text).
	    {\it Hanning}: The same spectrum after being smoothy by a Hanning filter (3 channels).
	}
  \label{o2_spec_final}
\end{figure*}

\begin{figure*}
  \resizebox{\hsize}{!}{ 
	\rotatebox{00}{\includegraphics{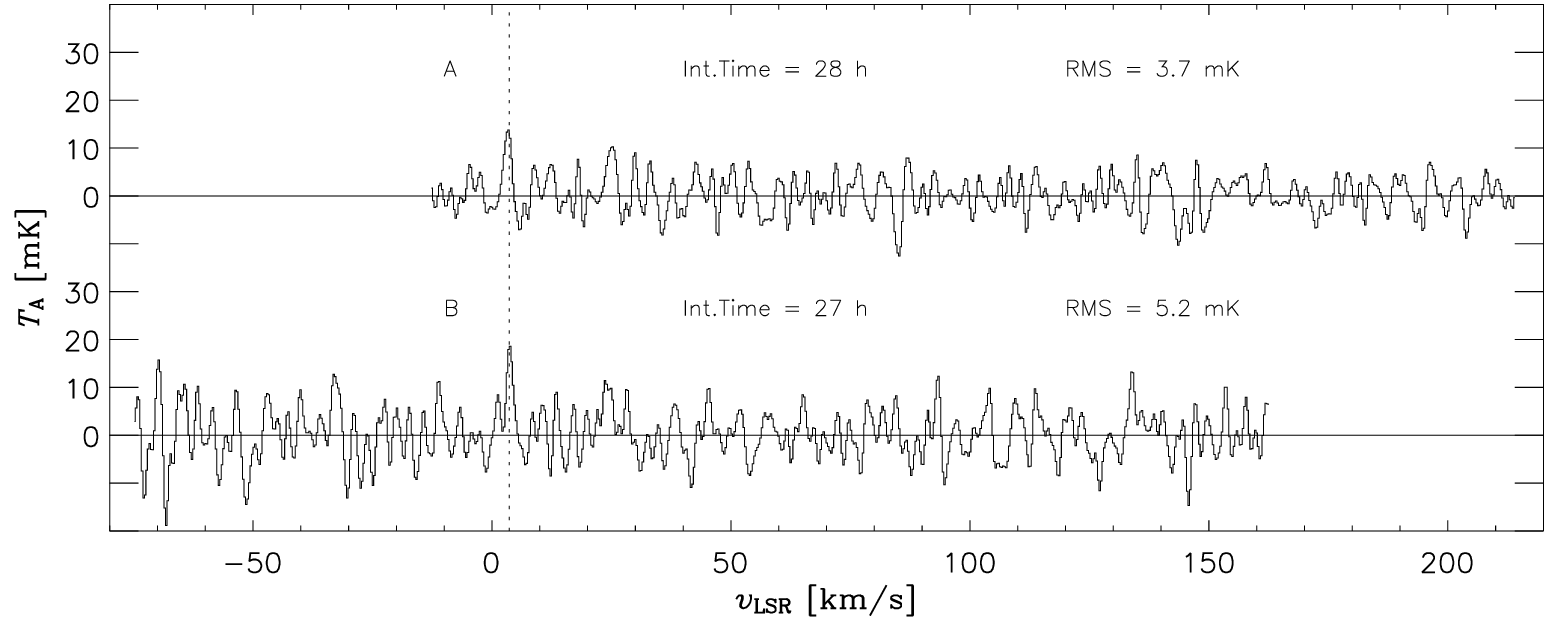}} 
	}
  \caption{ The finally reduced spectral data, which have been
		divided into two halves A and B (see lower panels in Fig.\,\ref{o2_reduce}),
 		showing the presence of the \molo\ feature also at these reduced observing times.
	}
  \label{o2_spec_parts}
\end{figure*}

\subsubsection{Noise behaviour and presentation of spectra}

In the left panel of Fig.\,\ref{o2_noise}, the rms noise of the finally selected data set is shown as a function of
the integration time. The 119\,GHz data follow the theoretical prediction for white noise reasonably well to the resolution
of the spectrometer, 292\,kHz or a little more then 2 channels (right panel Fig. \,\ref{o2_noise}).
As a final reduction step the spectrum was passed through a Hanning filter of a 3-channel width to accommodate
the resolution of the spectrometer. As seen in the right panel of Fig.\,\ref{o2_noise} the result of this was that
the effective resolution deteriorates to about 4 channels or 500\,kHz.

The finally reduced data are shown in Fig.\,\ref{o2_spec_final}.  The spectrum
reveals two line features clearly above the noise level. These are the \molo\ ($N_J=1_1 - 1_0)$ line and a feature that
coincides in frequency with c-C$_2$H$_4$O $(3_{2,\,2}-2_{1,\,1})$. In order to gauge the reality of these detections
the data set was divided into two halves (Figs.\,\ref{o2_reduce} and \ref{o2_spec_parts}).  Obviously, the \molo\
spectral feature is clearly seen in both spectra, lending further confidence to the reliability of the reduction
method.

\end{document}